\documentclass[prl,preprint,a4paper,floatfix,superscriptaddress]{revtex4-1}

\usepackage{graphicx}
\usepackage{amsmath}
\usepackage{hyperref}
\usepackage{url}
\usepackage{epstopdf}

\newcommand{\fig}{Fig.}

\begin{document}

\title{Infrared study of the spin reorientation transition and its reversal in the superconducting state in underdoped Ba$ _{1-x} $K$ _{x} $Fe$ _{2} $As$ _{2} $}

\author{B.P.P. Mallett}
\affiliation{University of Fribourg, Department of Physics and Fribourg Center for Nanomaterials, Chemin du Musée 3, CH-1700 Fribourg, Switzerland}

\author{P. Marsik}
\affiliation{University of Fribourg, Department of Physics and Fribourg Center for Nanomaterials, Chemin du Musée 3, CH-1700 Fribourg, Switzerland}

\author{M. Yazdi-Rizi}
\affiliation{University of Fribourg, Department of Physics and Fribourg Center for Nanomaterials, Chemin du Musée 3, CH-1700 Fribourg, Switzerland}

\author{Th. Wolf}
\affiliation{Institute of Solid State Physics, Karlsruhe Institute of Technology, Postfach 3640, Karlsruhe 76021, Germany}

\author{A.E. B\"{o}hmer}
\affiliation{Institute of Solid State Physics, Karlsruhe Institute of Technology, Postfach 3640, Karlsruhe 76021, Germany}

\author{F. Hardy}
\affiliation{Institute of Solid State Physics, Karlsruhe Institute of Technology, Postfach 3640, Karlsruhe 76021, Germany}

\author{C. Meingast}
\affiliation{Institute of Solid State Physics, Karlsruhe Institute of Technology, Postfach 3640, Karlsruhe 76021, Germany}

\author{D. Munzar}
\affiliation{Department of Condensed Matter Physics, Faculty of Science and Central European Institute of Technology, Masaryk University, Kotl\'{a}\v{r}ska 2, 61137 Brno, Czech Republic}

\author{C. Bernhard}
\affiliation{University of Fribourg, Department of Physics and Fribourg Center for Nanomaterials, Chemin du Musée 3, CH-1700 Fribourg, Switzerland}

\date{\today}

\pacs{ 74.70.−b, 74.25.Gz, 78.30.−j}
\keywords{}

\begin{abstract}

With infrared spectroscopy we investigated the spin-reorientation transition from an orthorhombic antiferromagnetic (o-AF) to a tetragonal AF (t-AF) phase and the reentrance of the o-AF phase in the superconducting state of underdoped Ba$ _{1-x} $K$ _{x} $Fe$ _{2} $As$ _{2} $. In agreement with the predicted transition from a single-$\mathbf{Q}$ to a double-$\mathbf{Q}$ AF structure, we found that a distinct spin density wave (SDW) develops in the t-AF phase.  The pair breaking peak of this SDW acquires much more low-energy spectral weight than the one in the o-AF state which indicates that it competes more strongly with superconductivity. We also observed additional phonon modes in the t-AF phase which likely arise from a Brillouin-zone folding that is induced by the double-$\mathbf{Q}$ magnetic structure with two Fe sublattices exhibiting different magnitudes of the magnetic moment.

\end{abstract}

\maketitle

Similar to the cuprate high-$ T_{c} $ superconductors (HTS), the phase diagram of the iron arsenides is characterized by the interaction of superconductivity (SC) with electronic correlations that give rise to competing orders and may also be involved in the SC pairing  \cite{paglione2010, stewart2011, chubukov2012, valenzuela2010, mazin2008,fernandes2014,medici2014}. The iron arsenides are considered to be archetypal HTS since they exhibit well resolved structural and magnetic phase transitions with long-range ordered states that, unlike the cuprates, are not obscured by strong fluctuation effects. However, the interpretation is still complicated by their multi-band and multi-orbital structure and a sizeable coupling between the charge, spin, orbital and lattice degrees of freedom \cite{paglione2010, stewart2011}. A prominent example is the combined structural and magnetic transition in undoped and lightly doped samples for which it is still debated whether it is driven by the spin or orbital sectors \cite{fernandes2014, lee2009,kruger2009}. A related, controversial issue is whether the antiferromagnetic (AF) order is better described by an itinerant picture, in terms of a spin density wave (SDW), or in terms of local moments with exchange interactions that depend on the orbital order.

New insights into these questions are expected from the recent observation of a spin-reorientation transition in underdoped Ba$ _{1-x} $Na$ _{x} $Fe$ _{2} $As$ _{2} $ \cite{avci2014,wasser2015} from the usual orthorhombic phase with a stripe-like AF  order (o-AF) at $ T^{\text{N}1} >T>T^{\text{N}2}$  to a new tetragonal phase with an out-of-plane spin orientation (t-AF) at $ T<T^{\text{N}2} $. Recently, in Ba$ _{1-x} $K$ _{x} $Fe$ _{2} $As$ _{2} $ even a reentrant transition back to the o-AF state has been observed at $ T^{\text{N}3} < T^{\text{N}2}$ that seems to be induced by SC \cite{bohmer2014}.

Different magnetic orders with a so-called double-$\mathbf{Q}$ structure, that can be viewed as a superposition of the single-$\mathbf{Q}$ waves with ordering vectors $ Q_{1} = (\pi, 0)$ and $ Q_{2}=(0, \pi) $ (the latter prevails in the o-AF phase), have been predicted for the t-AF phase within the magnetic \cite{kang2014, *wang2015, *fernandes2015} and the orbital scenario \cite{khalyavin2014}. The determination of the magnetic space group and the possibly related orbital order and/or structural modulations may be key to the identification of the underlying mechanism. Also of great interest is the effect of these magnetic states on the charge carrier dynamics and their relationship with SC.

Here we present a study of the in-plane infrared (IR) response of a Ba$ _{1-x} $K$ _{x} $Fe$ _{2} $As$ _{2} $ single crystal which exhibits the above described structural and magnetic transitions, including a reentrant behavior in the SC state.

\begin{figure}
		\includegraphics[width=0.80\columnwidth]{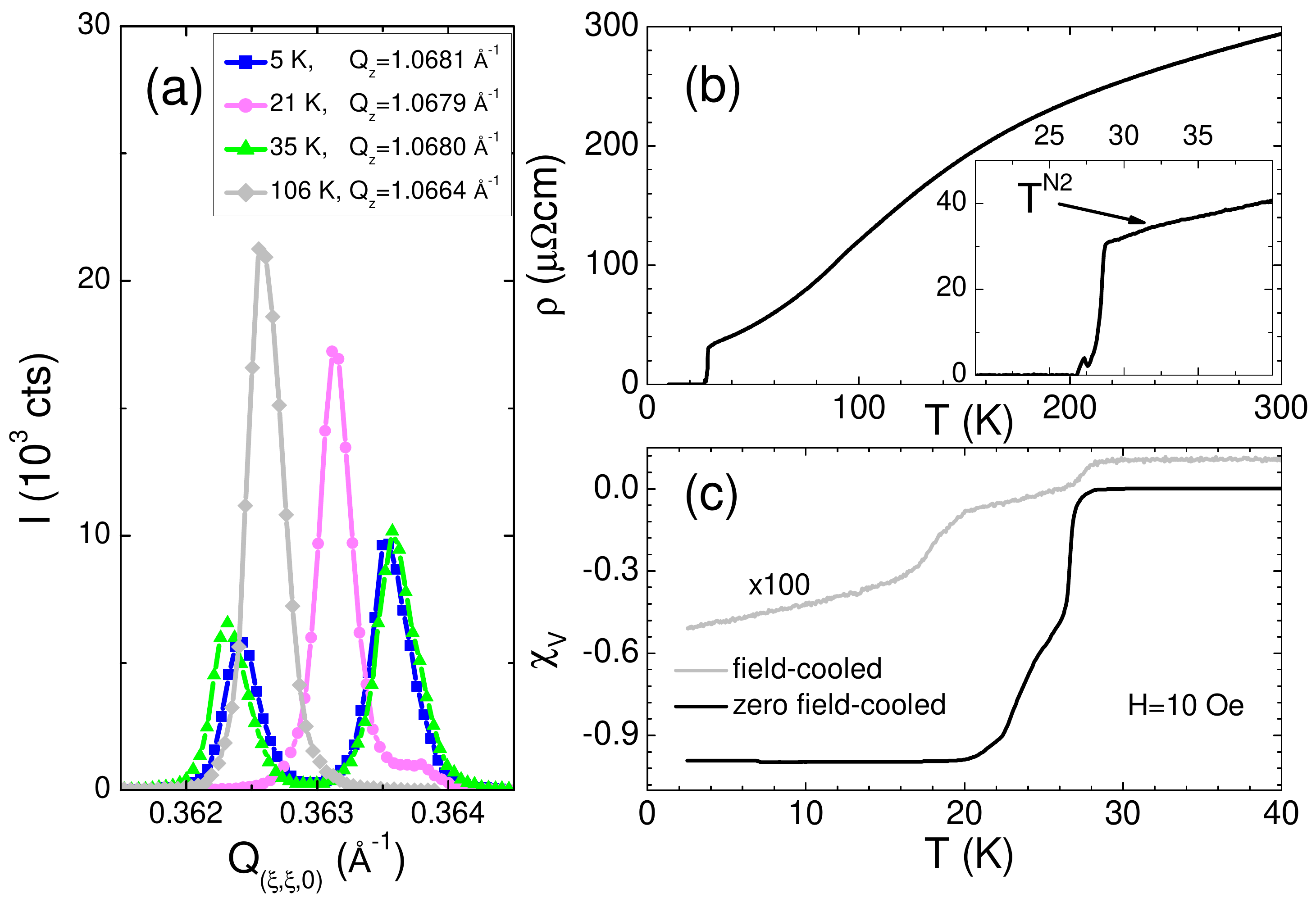}
	\caption{\label{fig1}(a) X-ray diffraction curves of the $ (1,1,14) $ Bragg peak showing the splitting in the o-AF and its absence in the t-AF states. (b) The dc resistivity, $ \rho $, with the inset detailing the changes around $ T_{c} $ and $ T^{\text{N}2}  $. (c) The volume susceptibility.}
\end{figure}

A crystal of  $ 2 \times 4 $~mm$ ^{2} $ was grown in an alumina crucible using an FeAs flux as described in Ref~\cite{karkin2014}. Crystals from the same batch were previously studied with thermal expansion and specific heat \cite{bohmer2014}. The $ c $-axis orientation of the spins in the t-AF state has been observed with neutron diffraction \cite{suprivate}. The K-content of one piece was determined by x-ray diffraction refinement to be $ x=0.247(2) $. The K-content of the present crystal has not been directly determined, but with muon spin rotation ($ \mu $SR) we identified three bulk magnetic transitions at $ T^{\text{N}1}\approx 72  $~K, $ T^{\text{N}2}\approx 32  $~K and $ T^{\text{N}3}\approx 18  $~K \cite{mallettbkfamusr2015} that conform with a K-content of $ x\approx 0.24- 0.25 $ \cite{bohmer2014}. 
The corresponding structural transitions are also seen in the x-ray diffraction data obtained with a 4-cycle diffractometer (RIGAKU Smartlab) equipped with a He-flow cryostat. As shown in Fig.~\ref{fig1}(a), the $ T $-evolution of the $ (1,1,14) $ Bragg peak confirms that in the t-AF phase the orthorhombic splitting is absent in about 90\% of the volume. Below $ T^{\text{N}3} $ the orthorhombic splitting reappears in the entire sample with a magnitude that is only slightly smaller than above $ T^{\text{N}2} $. The resistivity and magnetization data as shown in Figs.~\ref{fig1}(b) and \ref{fig1}(c) were obtained with a PPMS.

 \begin{figure*}

 		\includegraphics[width=1.0\textwidth]{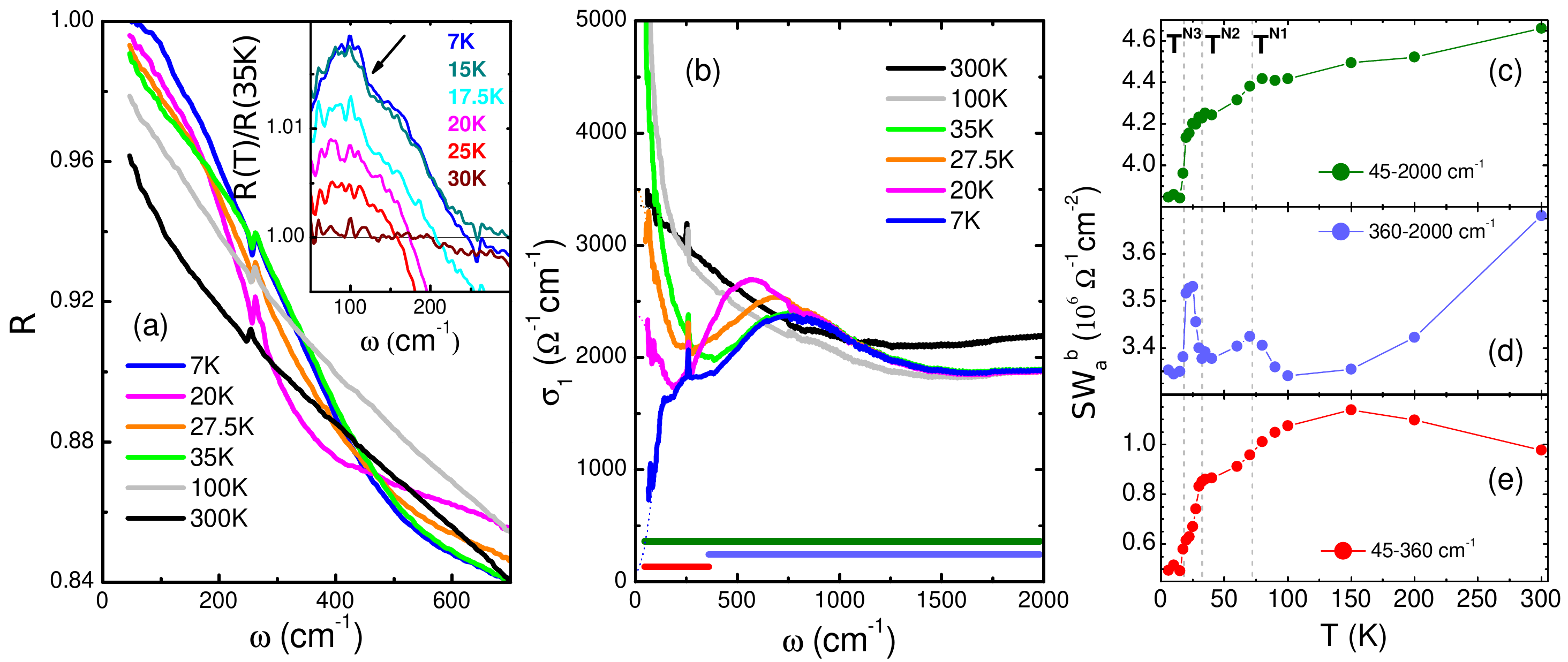}
 	\caption{\label{fig2}(a) The FIR reflectivity, $ R(\omega) $, at selected temperatures. Inset; spectra of the ratio $ R(T)/R(T=35$~K$) $. Note the additional gap feature at 130~cm$ ^{-1} $ which develops below 20~K.  (b) The corresponding spectra of the real part of the optical conductivity, $ \sigma_1 $. (c-e) The integrated spectral weight, $ SW_{a}^{b}=\int_{a}^{b}\sigma_{1}(\omega)d\,\omega $, for the relevant frequency intervals.}
 	
 \end{figure*}

The reflectivity, $ R $, was measured at $ 45-700 $~cm$ ^{-1} $ with a Bruker Vertex 70v FTIR spectrometer with an in situ gold evaporation technique \cite{homes1993,kim2010infrared}. For the ellipsometry measurements we used a home built rotating-analyzer setup attached to a Bruker 113v at $ 200-4500 $~cm$ ^{-1} $ \cite{bernhard2004thinfilms} and a Woollam VASE ellipsometer at $ 4000-52000 $~cm$ ^{-1} $. We measured the in-plane component of the optical response, the twinning of the sample in the o-AF state was not controlled. The combined ellipsometry and reflectivity data have been analyzed as described in Refs.~\cite{kim2010infrared, kuzmenko2005} to obtain the complex optical response functions.

Representative spectra of $ R(\omega) $ are shown in Fig.~\ref{fig2}(a), for the corresponding spectra of the real part of the optical conductivity, $ \sigma_{1} $, see \fig~\ref{fig2}(b). In the paramagnetic state at $ T>T^{\text{N}1} $ and in the o-AF state at $ T^{\text{N}1}>T>T^{\text{N}2} $ the spectra are similar as previously reported for underdoped and undoped Ba$ _{1-x} $K$ _{x} $Fe$ _{2} $As$ _{2} $ \cite{hu2008,schafgans2011,nakajima2011,marsik2013, dai2012}. Above $ T^{\text{N}1} $, they display a Drude peak at low frequencies due to the itinerant carriers, and a pronounced tail to higher frequencies with an upturn toward a maximum around 5000~cm$ ^{-1} $, that arise from inelastic scattering and low-energy interband transitions, respectively. With decreasing $ T $ the Drude peak narrows and the conductivity decreases overall due to a spectral weight transfer to high energies that originates from the strong Hund's-rule-coupling \cite{schafgans2011,wang2012,georges2013}.

The well-known features due to the SDW in the o-AF state are a strong spectral weight (SW) reduction and narrowing of the Drude-peak, a partial, gap-like suppression of $ \sigma_{1} $ between 100 and 500~cm$ ^{-1} $ and the formation of a band around 750~cm$ ^{-1} $ that corresponds to the SDW pair-breaking peak \cite{hu2008,schafgans2011,nakajima2011,marsik2013}. These characteristic features also manifest themselves in the $ T $-dependence of the integrated SW which is displayed in Figs.~\ref{fig2}(c-d) for representative frequency ranges. Figure~\ref{fig2}(c) shows the overall SW decrease due to the Hund's-rule-coupling. Figure~\ref{fig2}(d) reveals that the SW change due to the SDW formation is rather gradual and starts well above $ T^{\text{N}1}\approx 72$~K. This may be due to precursor fluctuations below the frequency probed by the experiment (THz in the IR compared with MHz in $ \mu $SR) that were previously also observed in underdoped Ba(Fe$ _{1-x} $Co$ _{x} $)$ _{2} $As$ _{2} $ \cite{marsik2013} and predicted \cite{mazin2008}. 

The SW of the pair breaking peak, which reflects the magnitude of the order parameter of the SDW and thus of the ordered magnetic moment due to the itinerant carriers,  amounts to $ 4.6\times10^5  $~$ \Omega^{-1} $cm$^{-2}$ \cite{[{Supplementary Information: }] malletttwox1342IR2015suppmat}. It is only about 40\% lower than in the undoped parent compound \cite{hu2008,marsik2010}. Such a moderate reduction of the magnetic moment is corroborated by a $ \mu $SR study \cite{wiesenmayer2011}. 

Next, we discuss the additional SW changes in the t-AF state below $ T^{\text{N}2}\approx 32 $~K. Figure~\ref{fig2} reveals that the transfer of SW from the Drude-peak to the SDW is now further enhanced. The change of the shape of the pair-breaking peak is rather specific. The high energy part remains almost unchanged whereas a large amount of additional SW is accumulated in a band centered at a lower energy.  This behaviour is suggestive of the scenario discussed in Ref.~\cite{eremin2010}, in which the SDW in the double-$ \mathbf{Q} $ state involves an additional set of electron- and hole-like bands that was not gapped in the o-AF state. The different orbital character of these bands may also be responsible for the spin reorientation toward the $ c $-axis that is seen in neutron diffraction \cite{suprivate}. As detailed in \cite{[{Supplementary Information: }] malletttwox1342IR2015suppmat}, the SW of the pair-breaking peak nearly doubles at $ 9.1\times10^5  $~$ \Omega^{-1} $cm$^{-2}$ and becomes even slightly larger (by about 20\%) than in the parent compound. In the future it will be interesting to compare these results with neutron scattering data, which equally probe the localized magnetic moment, to obtain an estimate of the relative magnitudes of the itinerant and localized moments.

\begin{figure*}

		\includegraphics[width=1.0\textwidth]{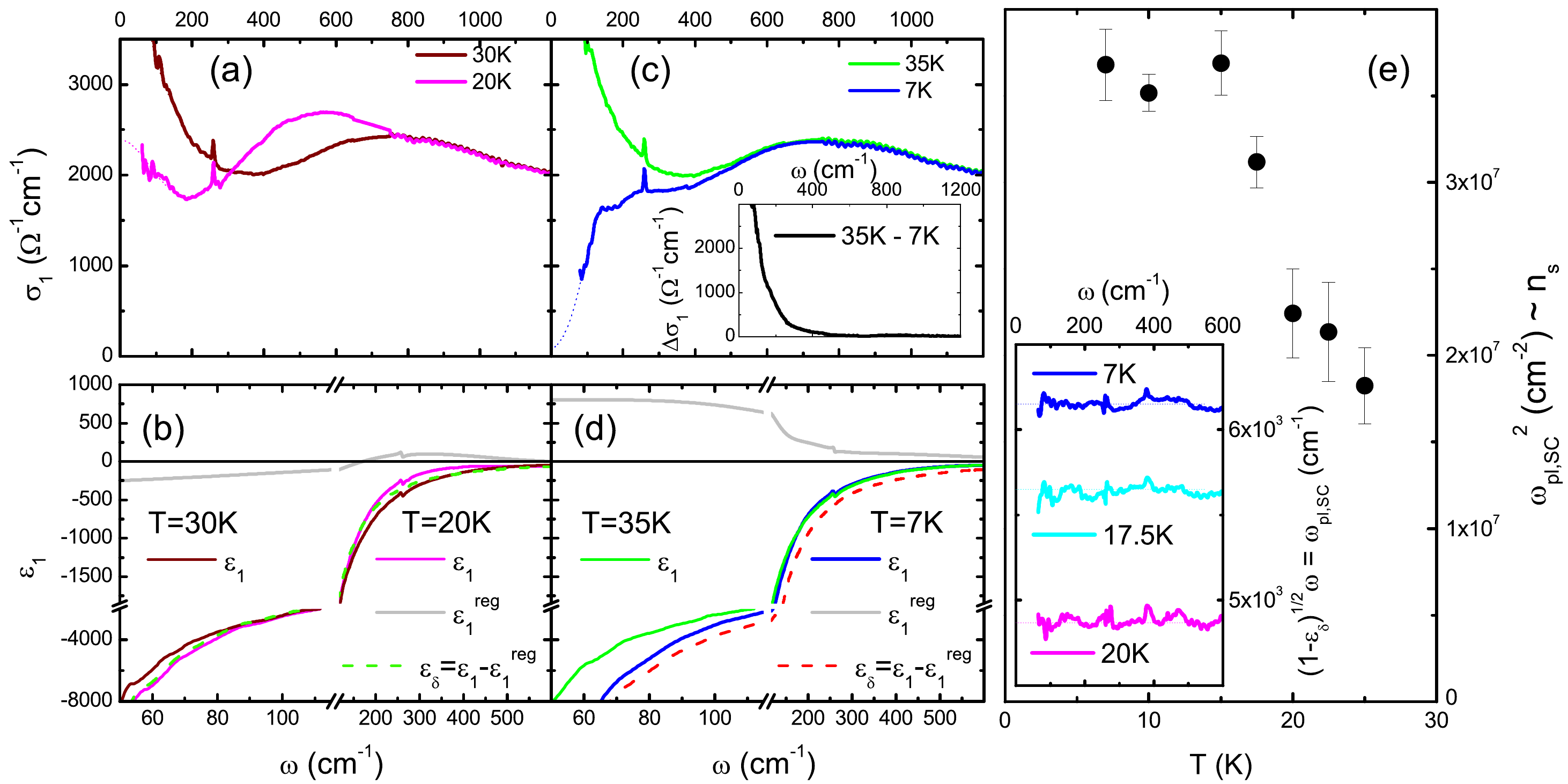}
	\caption{\label{fig3} FIR spectra of (a) $ \sigma_{1} $ and (b) $ \epsilon_{1} $ in the t-AF phase above and below $ T_{\mathrm c}^{\mathrm {ons}}=28 $~K. Also shown in (b) are the derived spectra of $ \epsilon_{1}^{\text{reg}} $ and $ \epsilon_{\delta} $ as discussed in the text and Ref.~\cite{[{Supplementary Information: }] malletttwox1342IR2015suppmat}. Panels (c) and (d) show an equivalent analysis in the o-AF phase with the inset to (c) showing the difference spectra $ \sigma_{1}(T=35$K$)-\sigma_{1}(T=7$K$) $.  (e) $ T $-dependence of $ \omega_{\text{pl,SC}}^{2} $ which has been derived from $ \epsilon_{\delta} $ as shown in the inset.}
	
\end{figure*} 

The large, additional SW of the SDW pair-breaking peak and the corresponding larger suppression of the Drude-response suggest that the t-AF state competes much more severely with SC than the o-AF state. In Ref.~\cite{bohmer2014}, SC was even reported to be absent above $ T^{\text{N}3} $ for a similar crystal with $ x=0.247 $. For our crystal we observe a weak SC response above $T^{\text{N}3} $ with an onset at $ T_{\mathrm c}^{\mathrm {ons}}=28 $~K in the resistive transition in \fig~\ref{fig1}(b) that is fairly broad and exhibits a peak feature that is reminiscent of a distribution of $ T_{c} $ values \cite{mosqueira1994}. The field-cooled magnetization in \fig~\ref{fig1}(c) exhibits a double transition with a step-like increase of the Meissner signal around $T^{\text{N}3} \approx 18$~K. Since the $ T_{c} $ values in the phase diagram of Ref.~\cite{bohmer2014} change significantly within a very narrow doping range and our x-ray data in Fig.~\ref{fig1}(a) indicate that a minor fraction of about 10\% remains in the o-AF phase, it is possible that the weak SC response above $ T^{\text{N}3} $ arises from the o-AF minority phase whereas the t-AF phase itself is not superconducting.

The FIR spectrum at 20 K confirms that the SC response at $ T^{\text{N}2} >T > T^{\text{N}3} $ is rather weak. The SC gap formation in the spectra of $ \sigma_{1}(\omega) $ remains incomplete and our estimate of the SC condensate density, $ n_{s} $, yields a rather small value. An upper limit for the SW of the delta function at the origin due to the SC condensate has been obtained, as shown in \fig~\ref{fig3}(a) and \ref{fig3}(b), by subtracting from $ \epsilon_{1}(\omega, 20$~K$) $ the contribution of the regular part, $ \epsilon_{1}^{\text{reg}} $, as derived from a Kramers-Kronig transformation of $ \epsilon_{2}(\omega>0, 20 $~K$)$ \cite{[{Supplementary Information: }] malletttwox1342IR2015suppmat}. The obtained value of the SC plasma frequency of $ \omega_{\text{pl,SC}}\approx 4800 $~cm$ ^{-1} $ corresponds to a magnetic penetration depth of $ \lambda_{ab}\approx 335 $~nm. Note that this analysis is based on the assumption of a homogeneous SC state which is not justified if only the o-AF minority phase is superconducting. On the other hand, it provides an upper limit for the SC condensate of the t-AF phase in case it is intrinsically superconducting.

Below the reentrant transition at $ T^{\text{N}3}\approx 18 $~K the spectra undergo characteristic changes which reveal that the SDW reverts to almost the same state as in the o-AF phase above $ T^{\text{N}2} $ and, concomitantly, the SC response is strongly enhanced.  
Figure~\ref{fig3}(c) shows that the pair-breaking peak of the SDW is virtually the same for the $ \sigma_{1}(\omega) $ spectra at 7~K and 35~K. Figure~\ref{fig2}(d) confirms that the SW which the pair-breaking peak has gained below $ T^{\text{N}2} $ is lost again below $ T^{\text{N}3} $ where it is redistributed to the delta function at the origin due to the SC condensate (similar as theoretically described in Ref.~\cite{fernandes2010}).  The difference between the $ \sigma_{1}(\omega) $ spectra at 35 and 7~K in Fig.~\ref{fig3}(c) reveals a clear gap-like suppression with a gradual onset around 500~cm$ ^{-1} $ and a sharp, step-like decrease below 130~cm$ ^{-1} $. The latter gap feature is also evident in the reflectivity ratio spectra in the inset of \fig~\ref{fig2}(a) in terms of a kink that develops below $ T^{\text{N}3}\approx18 $~K.  
The sudden, additional increase of $ n_{s} $ below $ T^{\text{N}3} $ is demonstrated in Fig.~\ref{fig3}(e) in terms of the $ T $-dependence of $ \omega_{pl,SC}^{2} \propto n_{s} $. The latter is consistently obtained from the analysis of the missing SW for the $ \sigma_{1}(\omega) $ spectra shown in Fig.~\ref{fig3}(c) and from the subtraction of the contribution of the regular response from the measured $ \epsilon_{1}(\omega) $ to be, at 7~K, $ \omega_{\text{pl,SC}} \approx 6100(100)$~cm$ ^{-1} $ and $ \lambda_{ab}\approx 265(5) $~nm.

This characteristic SW redistribution between the SDW pair-breaking peak and the SC condensate demonstrates that the t-AF state is a more severe competitor for SC than the o-AF state. The reentrant transition to the o-AF state below $T^{\text{N}3} $ is therefore likely driven by the competition with SC.  Even if it turns out that another magnetic or electronic interaction is causing this transition, the onset of SC certainly will further stabilize the o-AF order with respect to the t-AF one.

\begin{figure}

		\includegraphics[width=0.65\columnwidth]{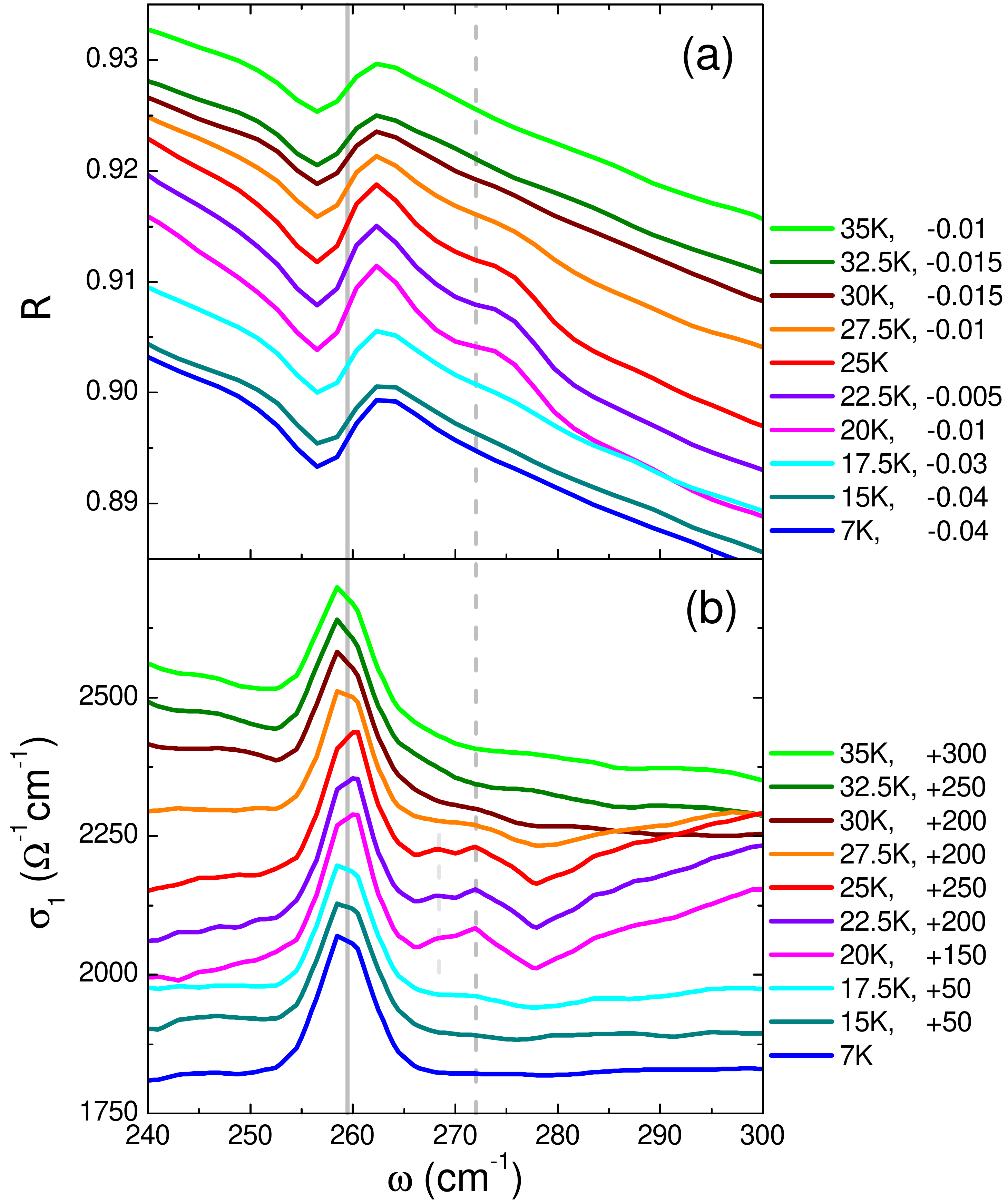}
	\caption{\label{fig4} (a) Reflectivity and (b) conductivity spectra (vertically offset as indicated in the legend) showing the additional IR-active phonon modes in the t-AF phase (marked by the dashed lines).}
	
\end{figure}

Finally, we discuss the $ T $-dependence of the IR-active mode around $ 259 $~cm$ ^{-1} $ which corresponds to the in-plane vibrations of Fe against As \cite{schafgans2011,litvinchuk2008,akrap2009}. Figure~\ref{fig4} shows that in the t-AF phase it is accompanied by a satellite at higher frequencies which consists of at least two peaks centered at 268 and 272~cm$ ^{-1} $. While their amplitudes are $ T $-dependent, the two frequencies are not. This suggests that the satellite is due to a Brillouin-zone (BZ) folding. The presence of the satellite is indeed consistent with the double-$ \mathbf{Q} $ magnetic structure with tetragonal symmetry as shown, for example, in Fig.~4 of Ref.~\cite{khalyavin2014}. It contains two types of Fe sites; Fe1 and Fe2 with zero and non-zero magnetic moment, respectively.  Due to the inequivalence of Fe1 and Fe2, the modes involving out-of-phase vibrations of Fe1 and Fe2 can be expected to be IR-active. A sketch of the corresponding displacement pattern is shown in Fig.~S4 of Ref.~\cite{[{Supplementary Information: }] malletttwox1342IR2015suppmat}. It reveals that the modes belong to the $ Z $-point of the BZ of the parent body-centered tetragonal lattice, presumably to the highest phonon branch. In the BZ of the t-AF phase, they are folded to the $ \Gamma $-point and thus can become IR-active and account for the satellite. This interpretation is consistent with the in-plane out-of-phase iron character of the highest branch along $ \Gamma $-$ Z $ in Ref.~\cite{sandoghchi2013}. The measured distance of ca. $ 13 $~cm$ ^{-1} $ between the main peak and the satellite is comparable to the calculated value of 18~cm$ ^{-1} $ between the highest $ \Gamma $-$ Z $ branch and the highest IR-active mode \cite{boeri2010}. The surprisingly large spectral weight of the satellite as compared to that of the main mode implies a considerable difference between the dynamical effective charges of Fe1 and Fe2. It remains to be seen whether it is accompanied by a corresponding static charge order and possibly local displacements that might be detected with x-ray diffraction.

In summary, with IR spectroscopy we studied the transition from an orthorhombic in plane antiferromagnetic (o-AF) to a tetragonal out-of-plane AF (t-AF) phase and the reentrance of the o-AF phase in the superconducting state of an underdoped Ba$ _{1-x} $K$ _{x} $Fe$ _{2} $As$ _{2} $ single crystal. In agreement with the predicted single-$\mathbf{Q}$ to double-$\mathbf{Q}$ transition our data show that a distinct SDW with a more pronounced pair breaking peak develops in the t-AF state. This SDW gives rise to a much stronger reduction of the free carrier spectral weight and thus competes more severly with SC than the SDW in the o-AF state. In the t-AF state we also observed additional IR-active phonon modes which likely originate from a Brillouin-zone folding that is induced by the double-$\mathbf{Q}$ magnetic structure with two Fe sublattices exhibiting different magnitudes of the magnetic moment.

\begin{acknowledgments}
This work was supported by the Schweizerische Nationalfonds (SNF) through grant No. 200020-153660. Some measurements were performed at the IR beam line of the ANKA synchrotron at FZ Karlsruhe, D where we acknowledge the support of Y.L. Mathis and M. S\"{u}pfle. We thank Yixi Su for sharing with us his unpublished neutron diffraction data and acknowledge valuable discussions with Lilia Boeri and Yurii Pashkevich.
\end{acknowledgments}

\end{document}